\documentclass[authorversion,manuscript,screen,sigconf]{acmart}

\usepackage[utf8]{inputenc}

\setcopyright{none}
\acmConference[Author Preprint]{Author Preprint}{2022}{}
\acmBooktitle{Author Preprint}
\copyrightyear{2023}
\acmYear{2023}
\begin{document}

% \title{There's Latency in My Game: Latency Makes Aiming Harder}
% \title{There's Latency in My Game, And Late Warp Makes it Better}
% \title{There's Latency in My Game, And Late Warp is the Antidote}
\title{Toward Understanding Display Size for FPS Esports Aiming}

\renewcommand{\shorttitle}{Display Size and FPS Esports}

% 4 authors on one row
\settopmatter{authorsperrow=4}

\author{Josef Spjut}
\orcid{0000-0001-5483-7867}
% \email{jspjut@nvidia.com}
\affiliation{%
  \institution{NVIDIA}
  \country{USA}
}
\author{Arjun Madhusudan}
% \email{amadhus2@ncsu.edu}
\affiliation{%
  \institution{NCSU \and NVIDIA}
  \country{USA}
}
\author{Benjamin Watson}
% \email{bwatson@ncsu.edu}
\affiliation{%
  \institution{NCSU}
  \country{USA}
}
\author{Seth Schneider}
% \email{sschneider@nvidia.com}
% \affiliation{%
%   \institution{NVIDIA}
% %   \streetaddress{P.O. Box 1212}
% %   \city{Dublin}
% %   \state{Ohio}
% %   \postcode{43017-6221}
% }
\author{Ben Boudaoud}
\orcid{0000-0003-4195-0793}
% \email{bboudaoud@nvidia.com}
% \affiliation{%
%   \institution{NVIDIA}
% %   \streetaddress{P.O. Box 1212}
% %   \city{Dublin}
% %   \state{Ohio}
% %   \postcode{43017-6221}
% }
\author{Joohwan Kim}
% \email{sckim@nvidia.com}
\affiliation{%
  \institution{NVIDIA}
  \country{USA}
}
\renewcommand{\shortauthors}{Spjut et al.}

\begin{abstract}
Gamers use a variety of different display sizes, though for PC gaming in particular, monitors in the 24 to 27 inch size range have become most popular.
Particularly popular among many PC gamers, first person shooter (FPS) games represent a genre where hand-eye coordination is particularly central to the player's performance in game. 
In a carefully designed pair of experiments on FPS aiming, we compare player performance across a range of display sizes.
First, we compare 12.5 inch, 17.3 inch and 24 inch monitors on a multi-target elimination task.
Secondly, we highlight the differences between 24.5 inch and 27 inch displays with a small target experiment, specifically designed to amplify these small changes.
We find a small, but statistically significant improvement from the larger monitor sizes, which is likely a combined effect between monitor size, resolution, and the player's natural viewing distance.
\end{abstract}



% \keywords{datasets, neural networks, gaze detection, text tagging}

\maketitle

\section{Introduction}
First person shooters (FPS) are one of the most popular esports genres.
A key feature of FPS games, particularly when using a mouse as the input device, is the high level of skill required to quickly and accurately hit targets.
A variety of factors contribute to a player's ability to complete these tasks, including the graphics and display.
To this end, we present two experiments investigating screen size impacts on FPS aiming tasks.
These studies cover some of the space between a carefully controlled lab environment and natural usage patterns.
For example, most players do not strictly control viewing distance or room brightness, though these parameters are essential to well controlled scientific experimentation.
The primary contribution of this work is studying natural player behavior in terms of viewing distance and head motion in contrast to conditions where viewing distance is more carefully controlled.

We used FirstPersonScience~\cite{Spjut19FPSci,boudaoud2022fpsci} (FPSci) to implement both experiments, a tool purpose-built for these types of FPS aiming task experiments~\cite{spjut2019latency,kim2020latewarp}. 
This program is similar to commercially available aim trainers, but allows low level control of the game loop and experiment configuration.

\section{Related Work}
We leave a thorough review of related work to others\footnote{For a more thorough list of related work, see \cite{kim2022display}}, and instead provide only a focused overview of some of the most relevant topics.

Today, computer systems for esports games most often use a desktop monitor as their main display. 
These monitors sport a variety of specifications including size, refresh rate, latency, and resolution, while also giving the players freedom to customize other features such as brightness or contrast as they wish. 
This creates a dynamic environment that can affect gameplay in many ways.
Many display-related effects have been considered by previous studies on FPS gaming, such as refresh rate, latency and size~\cite{ivkovic2015quantifying,spjut2019latency,kim2022display,liu2023effects}.
Additionally, this work has expanded on these topics by studying both network and local latency in FPS games~\cite{liu2021comparing,liu2021network,liu2021local}.

The study most directly related to this work on display sizes for FPS games was completed concurrently by Kim et al.~\shortcite{kim2022display}.
They also use FirstPersonScience to study varying screen sizes from 13-65", but in contrast to our work, they maintain a constant display field of view by controlling the viewing distance of the user. 
Their main finding is that the smallest screen size (13") demonstrates a statistically significant disadvantage to aiming performance when compared to each of the larger display sizes, with limited additional significant effects for displays larger than the 26" size. 

Other work has studied display size, but outside the context of FPS gaming.
A few studies report that display size has a variety of positive effects resulting in improved reaction times~\cite{tan2006physically,ni2006increased,hancock2015effects,browning2014screen} and greater immersion~\cite{hou2012effects}.
% Hancock et al. found that display sizes 40\inch and larger reduced cognitive load and reaction time \shortcite{hancock2015effects}, though larger screens also had higher resolution. 
% Two other studies found that larger screens improved spatial task performance by facilitating egocentric spatial perception \cite{tan2006physically, ni2006increased}, which may be related to increased sense of immersion when using larger displays \cite{hou2012effects}. 
% Browning and Teather measured 2D targeting performance on four display sizes ranging from 3.9 to 15.6\inch and observed performance reduction in the smallest display size \cite{browning2014screen}, where resolution and FoV changed with display sizes. 
In contrast, other studies show that display size has little or even opposite effects at a variety of similar tasks~\cite{riecke2009display,spittle2010effect,wang2013exploring}
% Display size did not affect egocentric distance estimation \cite{riecke2009display} or perceptual decision-making times in traditional sports \cite{spittle2010effect}. 
% Wang et al. measured 2D targeting performance on four display sizes ranging from 10.6 to 55\inch with resolution and FoV kept constant, where performance degraded at larger display sizes\cite{wang2013exploring}. 
% Overall, prior work suggests that larger display size can improve performance for some tasks, but leaves us uncertain that display size has similar effects in PC gaming. 
Our work focuses on FPS gaming, which requires precise aim and high levels of aiming skill, and we hope to provide some clarity on this open question, at least for this type of game-specific task.

Broadening to aspects of gaming outside of display size, others have considered the effects that in-game settings have on player performance~\cite{madhusuda2021dota2}.
Some lines of investigation explore the impact that variable refresh rate (i.e., G-SYNC) has on player performance, finding narrow improvements at times~\cite{riahi2021playing,watson2019effects}.

%  The previously mentioned study by Spjut, et al.~\cite{spjut2019latency} looks into how performance varies with refresh rates. Another study talks about how players choose to customize the features available in game \cite{madhusuda2021dota2}. A study closely related to this one looks into display sizes ranging from 13" to 65" showcases the effect of screen sizes when held at a constant FoV (should we cite this?)

% FPS-related research methodologies have been gaining popularity. Recent studies dealing with latency use CG:GO, a widely played FPS game \cite{liu2021network} and \cite{liu2021local}. Not only is FPS one of the more popular modes amongst esports games, many of its core mechanics such as mouse movement, perception, reaction times, etc are also an integral part of esports in other genres as well.

\section{Experiment 1: Screen size variation}

This experiment tests the effects of varying target size (task difficulty) alongside screen size differences. 
Participants used a 24 inch monitor and FPSci to simulate 3 different screen sizes - 24 inch commonly used in desktop setups, 17.3 inch available on laptops, and 12.5 inch tablet display. 
The experiment used a 240 Hz monitor, which can be found by players. 

The task for this experiment consisted of the player pressing shift while aiming at a red centering target, after which 8 static (green) targets appeared at once.
Participants had to move the mouse to align the in-game crosshair with each target, and left click once to eliminate it.
A trial is completed once all 8 targets are eliminated.
Each target spawned in one of 8 separate regions, organized in a grid around the centering target, at a random location within their assigned region.
Each trial's 8 targets could be one of 4 different sizes, with all targets equally sized within a trial, where the smallest size was particularly challenging.

\begin{figure}
    \centering
    \includegraphics[width=\columnwidth]{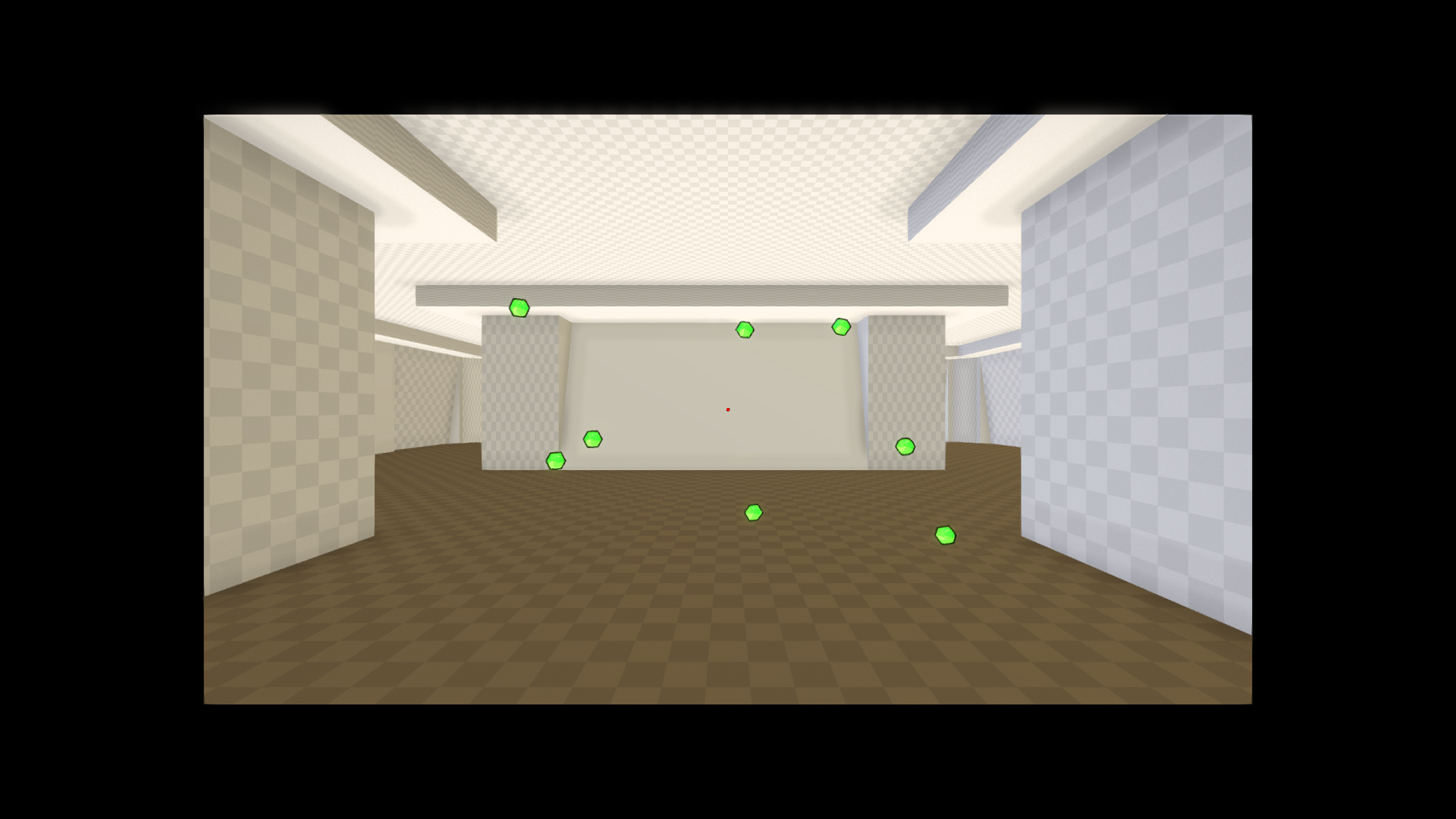}
    \caption{Experiment 1: Emulating a smaller screen with black borders on the 24 inch monitor used in our experiment.}
    \label{fig:smallscreen}
\end{figure}

The experiment consisted of 3 different display size sessions, where each session had 15 blocks of 4 trials. 
While each session was identical in task, the game was rendered on a different portion of the 24 inch monitor, simulating a smaller screen size. 
The region outside the rendered area was black with relevant content always shown in the middle of the screen (Fig.~\ref{fig:smallscreen}).

\subsection{Subjects}
14 participants completed this experiment in a counterbalanced ordering of display size sessions. 
All subjects were experienced FPS gamers.

\subsection{Results}
\begin{figure}
    \centering
    \includegraphics[width=\columnwidth]{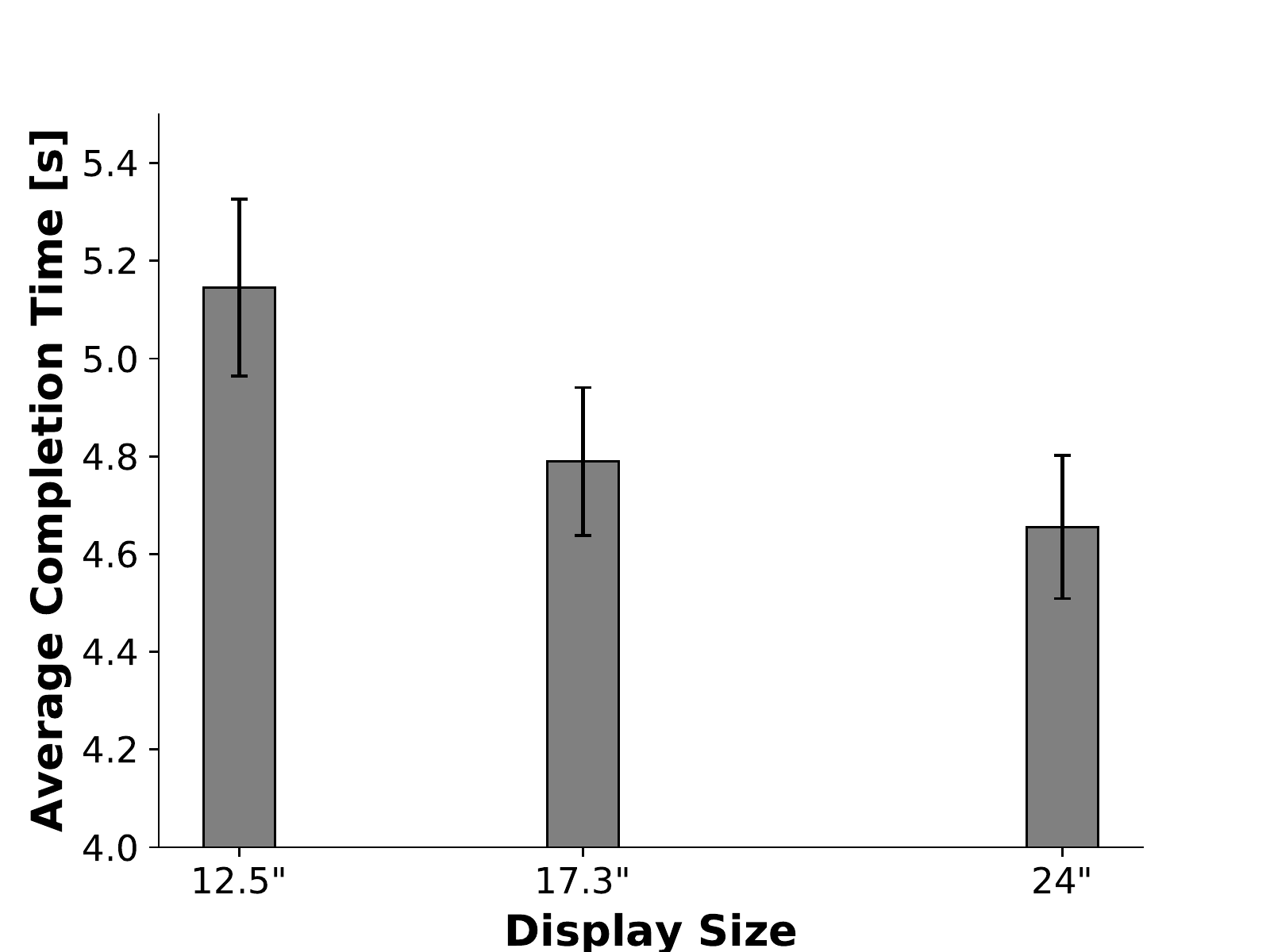}
    \caption{Plot demonstrating the effect of display size (in diagonal inches) on average task completion time across all 14 subjects in our study. Y error bars are standard error metric.}
    \label{fig:exp1_results}
\end{figure}

\begin{figure}
    \centering
    \includegraphics[width=\columnwidth]{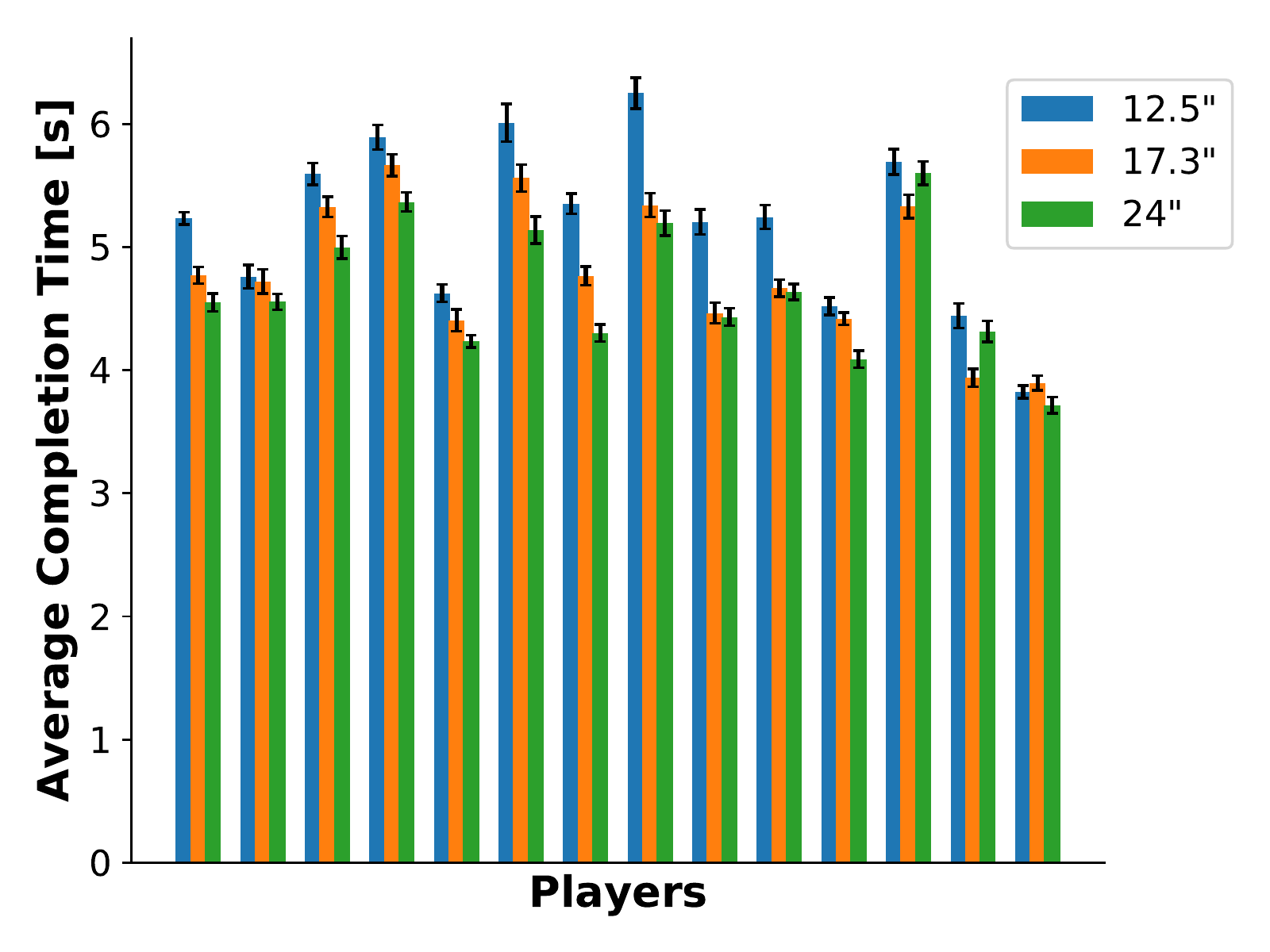}
    \caption{Experiment 1 results per subject by display size (in diagonal inches), note the almost ubiquitous trend of average task completion time scaling with display size. Error bars are standard error metric.}
    \label{fig:exp1_user_results}
\end{figure}

Participants not only had longer completion times for smaller screen sizes, they also demonstrated slower performance for the smaller targets.
This is visualized overall in Fig. \ref{fig:exp1_results} and per subject in Fig. \ref{fig:exp1_user_results}.

\subsection{Discussion}
A similar screen and target size experiment was also conducted with moving targets. 
While static targets often test the ‘flicking’ skills of a player, moving targets test another crucial aiming task that requires visual tracking. 
Similar trends of smaller screen and target sizes resulted in longer completion times.

\section{Experiment 2: Small Targets}

\begin{figure}
    \centering
    \includegraphics[width=\columnwidth]{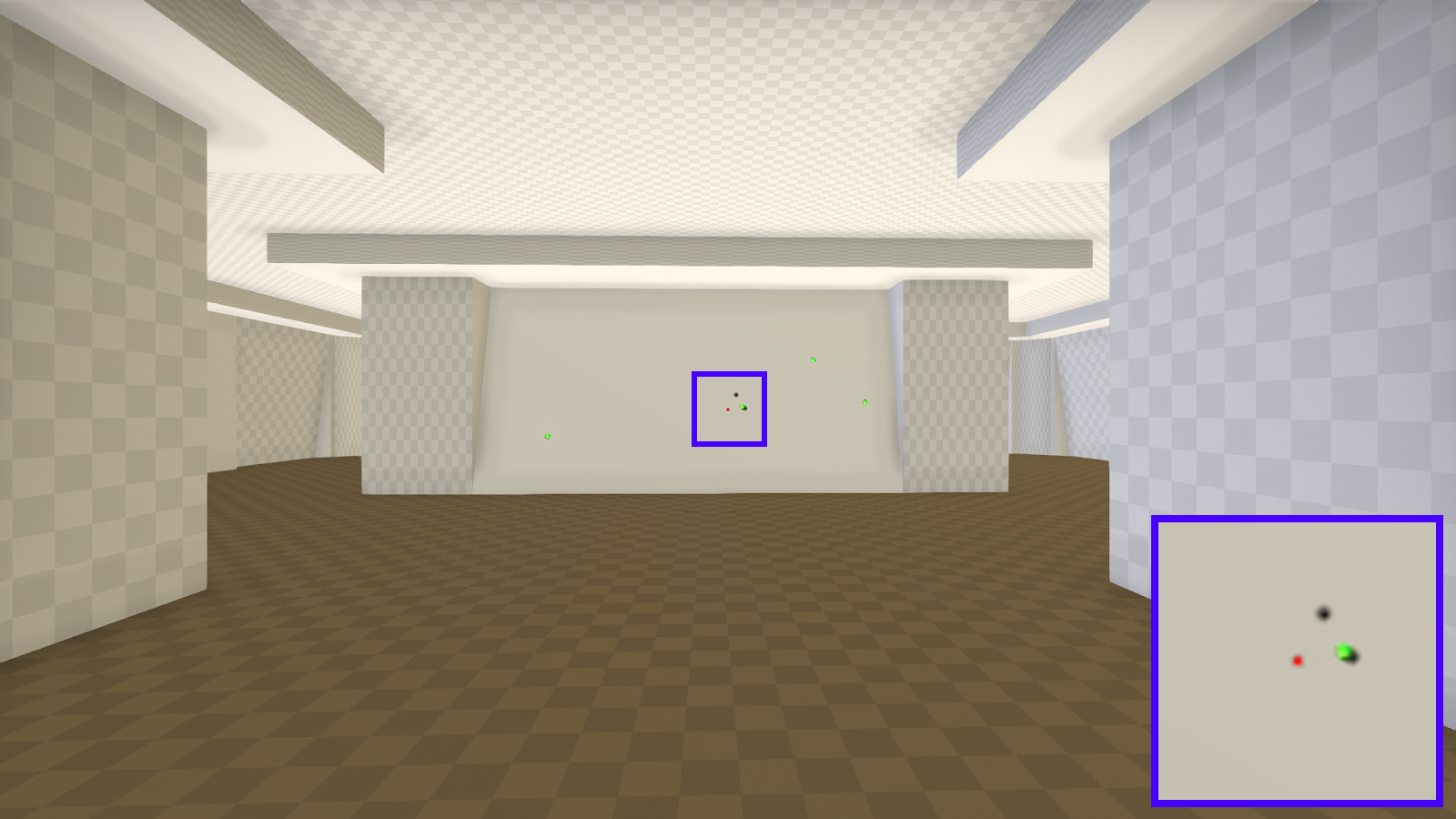}
    \caption{Experiment 2: small targets. Targets are intentionally small to amplify the effect of small changes in screen size (24.5 inch vs. 27 inch) and resolution (1920x1080 vs. 2560x1440).}
    \label{fig:smalltargets}
\end{figure}

The second experiment was designed to test the difference between 24.5 and 27 inch displays with resolutions of 1920x1080 (1080p) and 2560x1440 (1440p) respectively, and was reported on in part in a previous blog post~\cite{spjut22smalltargets}.
Given that esports gamers prefer high refresh rate displays to minimize latency and achieve maximum visual fluidity, both of these displays refresh at 360 Hz.
The task was specifically designed to try to amplify this small difference in screen size and display resolution, its defining feature being that the targets were very small on screen.
In fact, they may not even be visible in the screenshot in Figure~\ref{fig:smalltargets}.

Each trial had each user complete a series of target eliminations where a group of 4 targets all appear at the same time at the start of the task in random locations within the defined parameters.
Similarly to the previous experiment, the trial is considered complete once the player clicks on all 4 targets one time each. 
This makes task completion time or "aiming time" a negative metric, that is lower aiming time is better. 
As a secondary measure, we can consider the player's accuracy, or number of hits over total number of shots. 
Since it takes time to click, we would expect these measures to be related.
This specific experiment design is available to download on github~\footnote{\url{https://github.com/NVlabs/FPSciSmallTargets}}.

\subsection{Subjects}
We had 13 NVIDIA employees complete this task 75 times each (as 5 blocks of 15 trials) on each of 2 different monitors. 
The first was a 24.5" 1080p 360 Hz Alienware 25 and the second was a 27" 1440p 360 Hz Asus ROG Swift PG27AQN. 
It took roughly 20 minutes for each user to complete all 150 trials. 
Each participant had the option to decline or stop participating at any point if they chose. 
One of the participants had difficulty with the task, therefore we exclude that user's results from the analysis. 
The remaining 12 participants had counterbalanced display ordering.

\subsection{Results}
\begin{figure}
    \centering
    \includegraphics[width=\columnwidth]{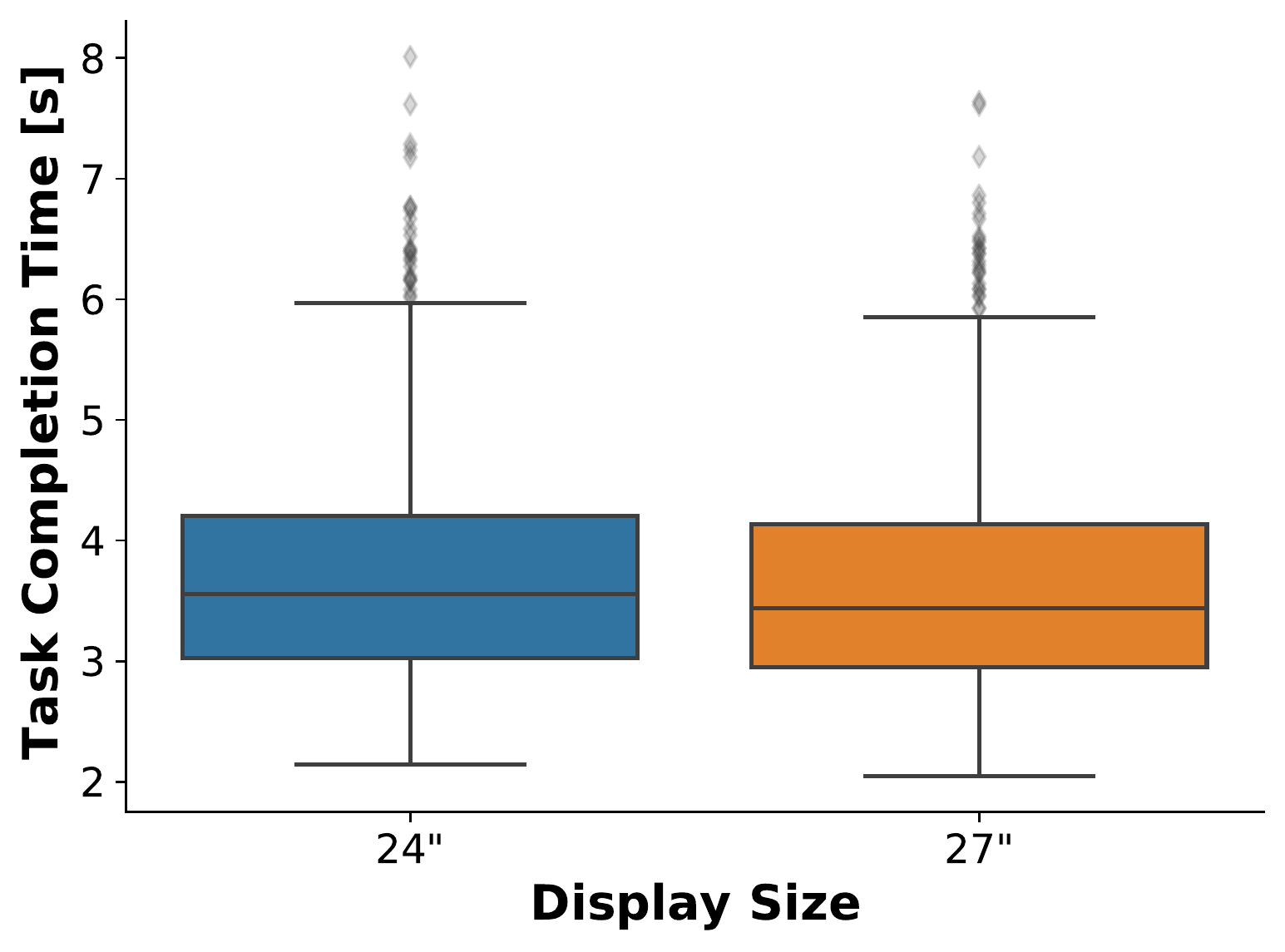}
    \caption{Box plot comparing the 24 and 27 inch conditions overall. Note that despite the small effect size, the difference in means reaches statistical significance in a paired t test.}
    \label{fig:ct_boxplot}
\end{figure}

The mean completion time for the 24.5" 1080p trials was 3.75 seconds while the 27" 1440p display resulted in a mean completion time of 3.64 seconds. 
The improvement in mean completion time for this experiment was therefore 111 ms, or 3\% of the total task time. 
We performed a pairwise t-test on this data indicating that the difference in these means is statistically significant for the 900 trials we included in the analysis (p=0.000325), see Fig.~\ref{fig:ct_boxplot}.
Pairwise t-tests on accuracy for the same users did not reach statistical significance.
Thus we conclude that the increase in speed of hitting targets did not come primarily from an increase in accuracy, but is instead an increase in aiming speed.

\begin{figure}
    \centering
    \includegraphics[width=\columnwidth]{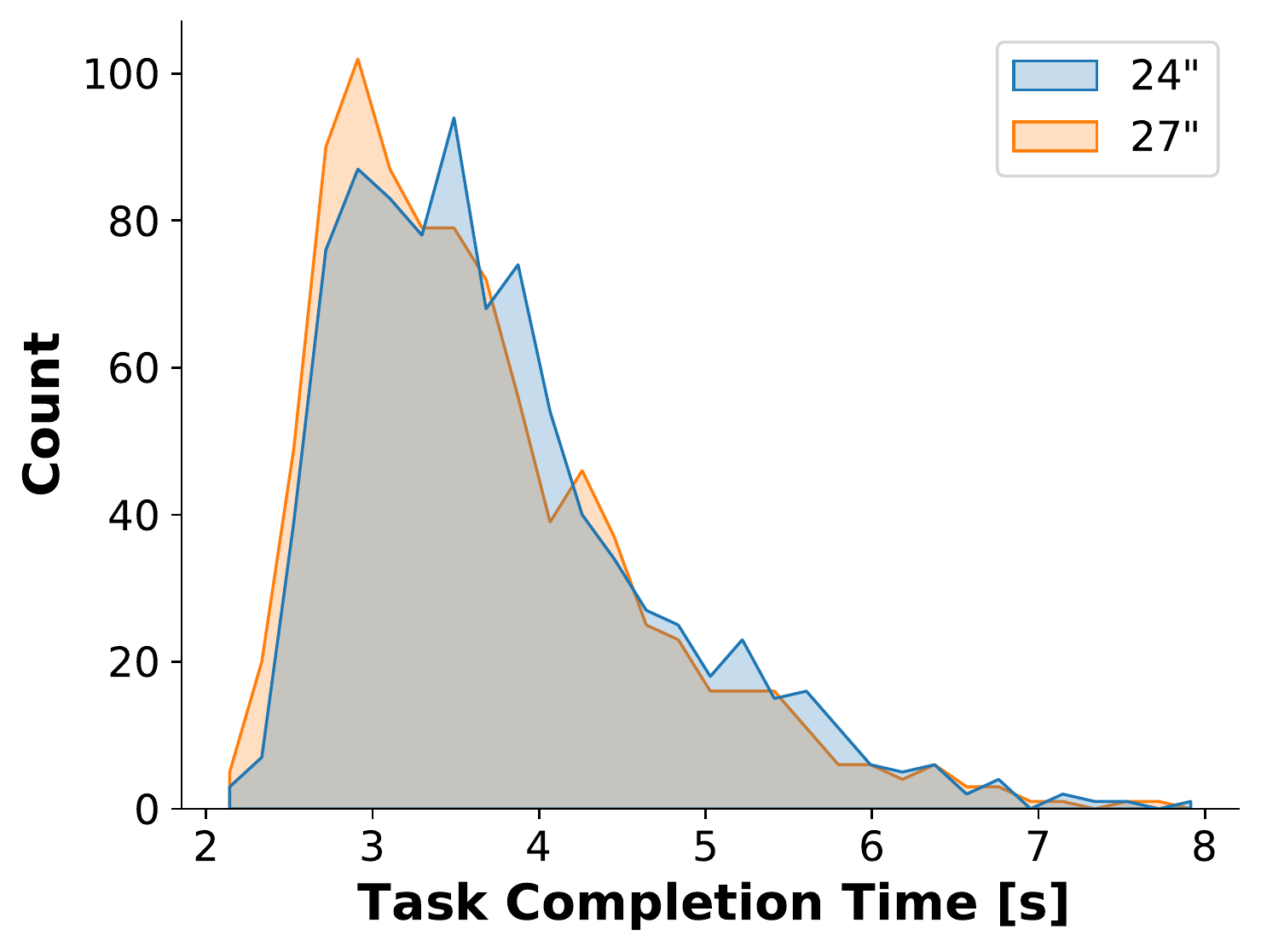}
    \caption{The experiment 2 task completion time histogram over all subjects shows a minor shift in mean and mode of completion times.}
    \label{fig:smalltargets_ct_hist}
\end{figure}

\begin{figure}
    \centering
    \includegraphics[width=\columnwidth]{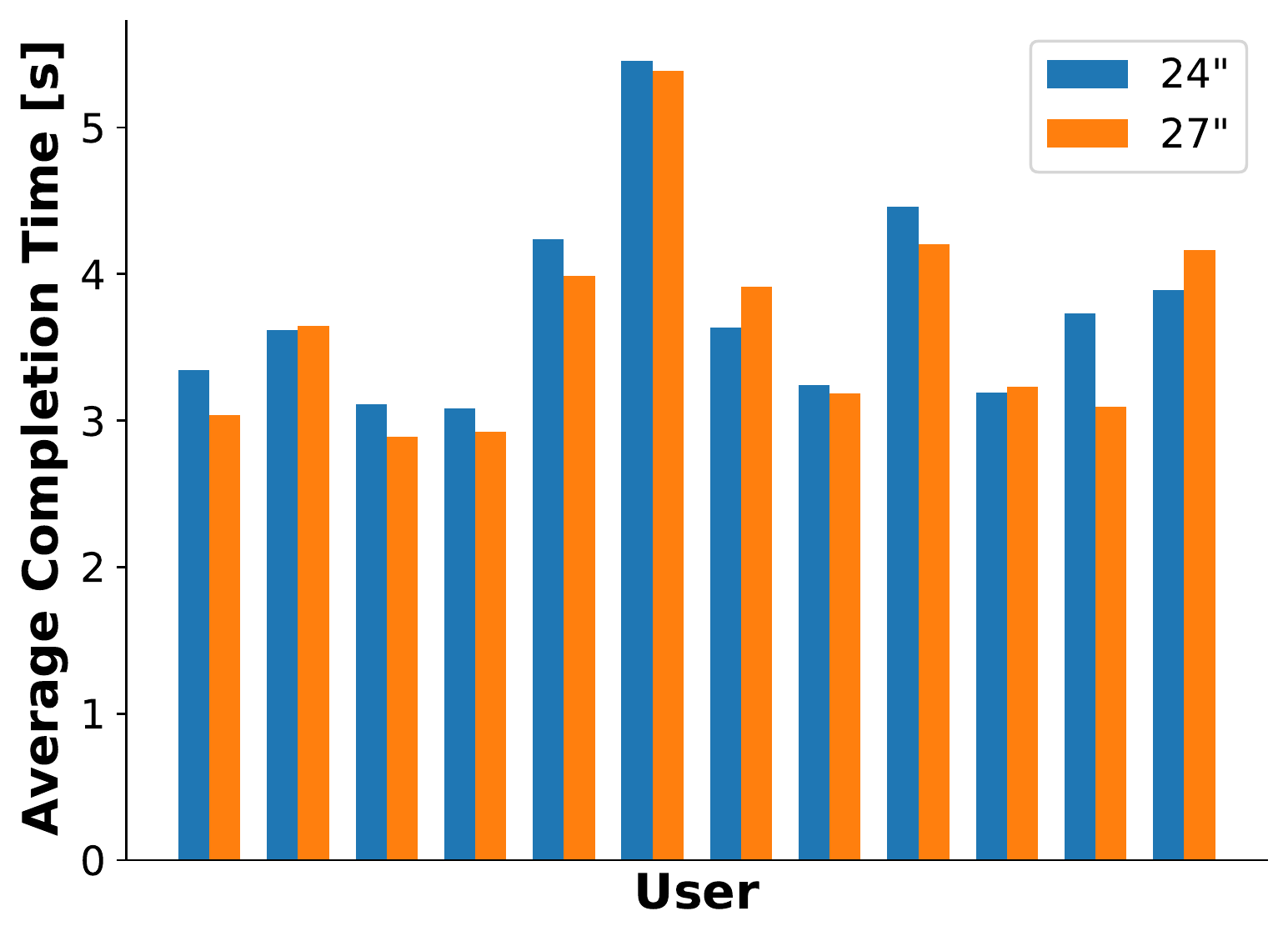}
    \caption{The experiment 2 per-subject average completion time showing no clear overall trend.}
    \label{fig:user_ct}
\end{figure}

The task completion time histogram (Fig.~\ref{fig:smalltargets_ct_hist}) reveals the minor differences in player behavior between the two screen sizes.
Slightly more trials were completed in a shorter time with the larger, higher resolution screen.
When we look at completion times per subject (Fig.~\ref{fig:user_ct}, the variation is much greater between subjects than between screen sizes.
This means that skilled users will almost certainly outplay novices even when the novice has the bigger display.
However, when players are evenly matched, then the display may still make this difference in some showdowns.

\section{Limitations and Future Work}
Since these experiments were intentionally designed to gather player performance results under variable display size conditions, allowing for a variety of subject behavior, there are some inherent limitations to our results.
First, viewing distances were not measured or controlled, a potentially clear confounding factor in the results.
Furthermore, the resolution of the in-game rendering was allowed to change.
Thus the results presented here should be considered as measurements of coupled factors, including display size, resolution, and player viewing distance.
Future work should strive to control one or more of these factors, and study the effects of specific changes.

Studying changes in resolution can be difficult as it is impossible to provide a higher or lower resolution image on a display of a given nominal resolution without some level of additional visual resampling artifacts being present.
If different displays are selected to allow for different nominal resolutions, this creates the possibility of additional confounding differences impacting results, such as brightness, contrast, color space, and subpixel arrangement.
Future work on display resolution should consider these factors.

Viewing distance and head rotation can, to an extent, be controlled with a chin rest or bite bar and could be measured with a variety of head tracking sensors.
As stated previously, we chose not to control this factor for these studies, though ideally we would have measured what naturally happened using head tracking.
A future study could focus on measuring and reporting changes in natural viewing distance arising from display size, which would shed more light on likely reasons for the performance changes observed in these studies.

Finally, display size changes can be quite significant with people regularly interacting with displays from small several inch screens on phones up to 80+ inch TVs and even theater-sized displays. 
Moreover, virtual reality headsets afford even wider field of view to users than are possible with most fixed displays. 
The design space available for screen size studies is so large that we recommend a more focused study that considers a reasonable range from laptop to TV sized displays, specifically in regard to FPS aiming.
A study coupling viewing distance to display size makes sense as changing display size may encourage different viewing distance, creating interesting effects in the field of view subtended by the displayed.

\section{Conclusions}

While neither of these experiments is fully conclusive on its own, this early evidence shows that increased screen size has some value for the most competitive of gamers.
There are many things players should train and optimize first, like game-specific and fundamental aiming skills, computer system latency, and CPU/GPU performance as well as the right balance of practice, diet and exercise~\cite{kari2019extended}.
Still, once all of the basics are in place, this work suggests that another small performance increase is possible by using a slightly larger, higher resolution display for FPS aiming tasks.
Furthermore, for tournament organizers looking to provide a fair experience for professional competitors, screen size should ideally be matched for all players.

\section*{Acknowledgements}

Doug Mendez assisted with administering the small targets experiment.

\bibliographystyle{ACM-Reference-Format}
\bibliography{main}


\begin{thebibliography}{23}

%%% the Web address.

\ifx \showCODEN    \undefined \def \showCODEN     #1{\unskip}     \fi
\ifx \showDOI      \undefined \def \showDOI       #1{#1}\fi
\ifx \showISBNx    \undefined \def \showISBNx     #1{\unskip}     \fi
\ifx \showISBNxiii \undefined \def \showISBNxiii  #1{\unskip}     \fi
\ifx \showISSN     \undefined \def \showISSN      #1{\unskip}     \fi
\ifx \showLCCN     \undefined \def \showLCCN      #1{\unskip}     \fi
\ifx \shownote     \undefined \def \shownote      #1{#1}          \fi
\ifx \showarticletitle \undefined \def \showarticletitle #1{#1}   \fi
\ifx \showURL      \undefined \def \showURL       {\relax}        \fi
% invisible to TeX
\providecommand\bibfield[2]{#2}
\providecommand\bibinfo[2]{#2}
\providecommand\natexlab[1]{#1}
\providecommand\showeprint[2][]{arXiv:#2}

\bibitem[\protect\citeauthoryear{Boudaoud, Spjut, and Kim}{Boudaoud
  et~al\mbox{.}}{2022}]%
        {boudaoud2022fpsci}
\bibfield{author}{\bibinfo{person}{Ben Boudaoud}, \bibinfo{person}{Josef
  Spjut}, {and} \bibinfo{person}{Joohwan Kim}.}
  \bibinfo{year}{2022}\natexlab{}.
\newblock \showarticletitle{FirstPersonScience: An Open Source Tool for
  Studying FPS Esports Aiming}.
\newblock In \bibinfo{booktitle}{\emph{ACM SIGGRAPH 2022 Talks}}.
  \bibinfo{pages}{1--2}.
\newblock


\bibitem[\protect\citeauthoryear{Browning and Teather}{Browning and
  Teather}{2014}]%
        {browning2014screen}
\bibfield{author}{\bibinfo{person}{Graeme Browning} {and}
  \bibinfo{person}{Robert~J. Teather}.} \bibinfo{year}{2014}\natexlab{}.
\newblock \showarticletitle{Screen scaling: Effects of screen scale on moving
  target selection}.
\newblock In \bibinfo{booktitle}{\emph{CHI'14 Extended Abstracts on Human
  Factors in Computing Systems}}. \bibinfo{pages}{2053--2058}.
\newblock


\bibitem[\protect\citeauthoryear{Hancock, Sawyer, and Stafford}{Hancock
  et~al\mbox{.}}{2015}]%
        {hancock2015effects}
\bibfield{author}{\bibinfo{person}{Peter~A. Hancock}, \bibinfo{person}{Ben~D.
  Sawyer}, {and} \bibinfo{person}{Shawn~C. Stafford}.}
  \bibinfo{year}{2015}\natexlab{}.
\newblock \showarticletitle{The effects of display size on performance}.
\newblock \bibinfo{journal}{\emph{Ergonomics}} \bibinfo{volume}{58},
  \bibinfo{number}{3} (\bibinfo{year}{2015}), \bibinfo{pages}{337--354}.
\newblock


\bibitem[\protect\citeauthoryear{Hou, Nam, Peng, and Lee}{Hou
  et~al\mbox{.}}{2012}]%
        {hou2012effects}
\bibfield{author}{\bibinfo{person}{Jinghui Hou}, \bibinfo{person}{Yujung Nam},
  \bibinfo{person}{Wei Peng}, {and} \bibinfo{person}{Kwan~Min Lee}.}
  \bibinfo{year}{2012}\natexlab{}.
\newblock \showarticletitle{Effects of screen size, viewing angle, and
  players’ immersion tendencies on game experience}.
\newblock \bibinfo{journal}{\emph{Computers in Human Behavior}}
  \bibinfo{volume}{28}, \bibinfo{number}{2} (\bibinfo{year}{2012}),
  \bibinfo{pages}{617--623}.
\newblock


\bibitem[\protect\citeauthoryear{Ivkovic, Stavness, Gutwin, and
  Sutcliffe}{Ivkovic et~al\mbox{.}}{2015}]%
        {ivkovic2015quantifying}
\bibfield{author}{\bibinfo{person}{Zenja Ivkovic}, \bibinfo{person}{Ian
  Stavness}, \bibinfo{person}{Carl Gutwin}, {and} \bibinfo{person}{Steven
  Sutcliffe}.} \bibinfo{year}{2015}\natexlab{}.
\newblock \showarticletitle{Quantifying and mitigating the negative effects of
  local latencies on aiming in 3d shooter games}. In
  \bibinfo{booktitle}{\emph{Proceedings of the 33rd Annual ACM Conference on
  Human Factors in Computing Systems}}. \bibinfo{pages}{135--144}.
\newblock


\bibitem[\protect\citeauthoryear{Kari, Siutila, and Karhulahti}{Kari
  et~al\mbox{.}}{2019}]%
        {kari2019extended}
\bibfield{author}{\bibinfo{person}{Tuomas Kari}, \bibinfo{person}{Miia
  Siutila}, {and} \bibinfo{person}{Veli-Matti Karhulahti}.}
  \bibinfo{year}{2019}\natexlab{}.
\newblock \showarticletitle{An extended study on training and physical exercise
  in esports}.
\newblock In \bibinfo{booktitle}{\emph{Exploring the cognitive, social,
  cultural, and psychological aspects of gaming and simulations}}.
  \bibinfo{publisher}{IGI Global}, \bibinfo{pages}{270--292}.
\newblock


\bibitem[\protect\citeauthoryear{Kim, Knowles, Spjut, Boudaoud, and
  Mcguire}{Kim et~al\mbox{.}}{2020}]%
        {kim2020latewarp}
\bibfield{author}{\bibinfo{person}{Joohwan Kim}, \bibinfo{person}{Pyarelal
  Knowles}, \bibinfo{person}{Josef Spjut}, \bibinfo{person}{Ben Boudaoud},
  {and} \bibinfo{person}{Morgan Mcguire}.} \bibinfo{year}{2020}\natexlab{}.
\newblock \showarticletitle{Post-Render Warp with Late Input Sampling Improves
  Aiming Under High Latency Conditions}.
\newblock \bibinfo{journal}{\emph{Proc. ACM Comput. Graph. Interact. Tech.}}
  \bibinfo{volume}{3}, \bibinfo{number}{2}, Article \bibinfo{articleno}{12}
  (\bibinfo{date}{aug} \bibinfo{year}{2020}), \bibinfo{numpages}{18}~pages.
\newblock
\urldef\tempurl%
\url{https://doi.org/10.1145/3406187}
\showDOI{\tempurl}


\bibitem[\protect\citeauthoryear{Kim, Madhusudan, Watson, Boudaoud, Tarrazo,
  and Spjut}{Kim et~al\mbox{.}}{2022}]%
        {kim2022display}
\bibfield{author}{\bibinfo{person}{Joohwan Kim}, \bibinfo{person}{Arjun
  Madhusudan}, \bibinfo{person}{Benjamin Watson}, \bibinfo{person}{Ben
  Boudaoud}, \bibinfo{person}{Roland Tarrazo}, {and} \bibinfo{person}{Josef
  Spjut}.} \bibinfo{year}{2022}\natexlab{}.
\newblock \showarticletitle{Display Size and Targeting Performance: Small
  Hurts, Large May Help}. In \bibinfo{booktitle}{\emph{SIGGRAPH Asia 2022
  Conference Papers}} (Daegu, Republic of Korea) \emph{(\bibinfo{series}{SA
  '22})}. \bibinfo{publisher}{Association for Computing Machinery},
  \bibinfo{address}{New York, NY, USA}.
\newblock
\showISBNx{9781450394703}
\urldef\tempurl%
\url{https://doi.org/10.1145/3550469.3555396}
\showDOI{\tempurl}


\bibitem[\protect\citeauthoryear{Liu, Claypool, Kuwahara, Scovell, and
  Sherman}{Liu et~al\mbox{.}}{2021a}]%
        {liu2021network}
\bibfield{author}{\bibinfo{person}{Shengmei Liu}, \bibinfo{person}{Mark
  Claypool}, \bibinfo{person}{Atsuo Kuwahara}, \bibinfo{person}{James Scovell},
  {and} \bibinfo{person}{Jamie Sherman}.} \bibinfo{year}{2021}\natexlab{a}.
\newblock \showarticletitle{The Effects of Network Latency on Competitive
  First-Person Shooter Game Players}. In \bibinfo{booktitle}{\emph{2021 13th
  International Conference on Quality of Multimedia Experience (QoMEX)}}.
  \bibinfo{pages}{151--156}.
\newblock
\urldef\tempurl%
\url{https://doi.org/10.1109/QoMEX51781.2021.9465419}
\showDOI{\tempurl}


\bibitem[\protect\citeauthoryear{Liu, Claypool, Kuwahara, Sherman, and
  Scovell}{Liu et~al\mbox{.}}{2021b}]%
        {liu2021local}
\bibfield{author}{\bibinfo{person}{Shengmei Liu}, \bibinfo{person}{Mark
  Claypool}, \bibinfo{person}{Atsuo Kuwahara}, \bibinfo{person}{Jamie Sherman},
  {and} \bibinfo{person}{James~J Scovell}.} \bibinfo{year}{2021}\natexlab{b}.
\newblock \showarticletitle{Lower is Better? The Effects of Local Latencies on
  Competitive First-Person Shooter Game Players}. In
  \bibinfo{booktitle}{\emph{Proceedings of the 2021 CHI Conference on Human
  Factors in Computing Systems}} (Yokohama, Japan) \emph{(\bibinfo{series}{CHI
  '21})}. \bibinfo{publisher}{Association for Computing Machinery},
  \bibinfo{address}{New York, NY, USA}, Article \bibinfo{articleno}{326},
  \bibinfo{numpages}{12}~pages.
\newblock
\showISBNx{9781450380966}
\urldef\tempurl%
\url{https://doi.org/10.1145/3411764.3445245}
\showDOI{\tempurl}


\bibitem[\protect\citeauthoryear{Liu, Kuwahara, Scovell, Sherman, and
  Claypool}{Liu et~al\mbox{.}}{2021c}]%
        {liu2021comparing}
\bibfield{author}{\bibinfo{person}{Shengmei Liu}, \bibinfo{person}{Atsuo
  Kuwahara}, \bibinfo{person}{James Scovell}, \bibinfo{person}{Jamie Sherman},
  {and} \bibinfo{person}{Mark Claypool}.} \bibinfo{year}{2021}\natexlab{c}.
\newblock \showarticletitle{Comparing the Effects of Network Latency versus
  Local Latency on Competitive First Person Shooter Game Players}.
\newblock  (\bibinfo{year}{2021}).
\newblock


\bibitem[\protect\citeauthoryear{Liu, Kuwahara, Scovell, and Claypool}{Liu
  et~al\mbox{.}}{2023}]%
        {liu2023effects}
\bibfield{author}{\bibinfo{person}{Shengmei Liu}, \bibinfo{person}{Atsuo
  Kuwahara}, \bibinfo{person}{James~J Scovell}, {and} \bibinfo{person}{Mark
  Claypool}.} \bibinfo{year}{2023}\natexlab{}.
\newblock \showarticletitle{The Effects of Frame Rate Variation on Game Player
  Quality of Experience}. In \bibinfo{booktitle}{\emph{Proceedings of the 2023
  CHI Conference on Human Factors in Computing Systems}}.
  \bibinfo{pages}{1--10}.
\newblock


\bibitem[\protect\citeauthoryear{Madhusudan and Watson}{Madhusudan and
  Watson}{2021}]%
        {madhusuda2021dota2}
\bibfield{author}{\bibinfo{person}{Arjun Madhusudan} {and}
  \bibinfo{person}{Benjamin Watson}.} \bibinfo{year}{2021}\natexlab{}.
\newblock \showarticletitle{Better Frame Rates or Better Visuals? An Early
  Report of Esports Player Practice in Dota 2}. In
  \bibinfo{booktitle}{\emph{Extended Abstracts of the 2021 Annual Symposium on
  Computer-Human Interaction in Play}} (Virtual Event, Austria)
  \emph{(\bibinfo{series}{CHI PLAY '21})}. \bibinfo{publisher}{Association for
  Computing Machinery}, \bibinfo{address}{New York, NY, USA},
  \bibinfo{pages}{174–178}.
\newblock
\showISBNx{9781450383561}
\urldef\tempurl%
\url{https://doi.org/10.1145/3450337.3483484}
\showDOI{\tempurl}


\bibitem[\protect\citeauthoryear{Ni, Bowman, and Chen}{Ni
  et~al\mbox{.}}{2006}]%
        {ni2006increased}
\bibfield{author}{\bibinfo{person}{Tao Ni}, \bibinfo{person}{Doug~A. Bowman},
  {and} \bibinfo{person}{Jian Chen}.} \bibinfo{year}{2006}\natexlab{}.
\newblock \showarticletitle{Increased display size and resolution improve task
  performance in information-rich virtual environments}. In
  \bibinfo{booktitle}{\emph{Proceedings of Graphics Interface 2006}}. Citeseer,
  \bibinfo{pages}{139--146}.
\newblock


\bibitem[\protect\citeauthoryear{Riahi and Watson}{Riahi and Watson}{2021}]%
        {riahi2021playing}
\bibfield{author}{\bibinfo{person}{Maryam Riahi} {and}
  \bibinfo{person}{Benjamin~Allen Watson}.} \bibinfo{year}{2021}\natexlab{}.
\newblock \showarticletitle{Am I Playing Better Now? The Effects of G-SYNC in
  60Hz Gameplay}.
\newblock \bibinfo{journal}{\emph{Proceedings of the ACM on Computer Graphics
  and Interactive Techniques}} \bibinfo{volume}{4}, \bibinfo{number}{1}
  (\bibinfo{year}{2021}), \bibinfo{pages}{1--17}.
\newblock


\bibitem[\protect\citeauthoryear{Riecke, Behbahani, and Shaw}{Riecke
  et~al\mbox{.}}{2009}]%
        {riecke2009display}
\bibfield{author}{\bibinfo{person}{Bernhard~E. Riecke},
  \bibinfo{person}{Pooya~Amini Behbahani}, {and} \bibinfo{person}{Chris~D.
  Shaw}.} \bibinfo{year}{2009}\natexlab{}.
\newblock \showarticletitle{Display size does not affect egocentric distance
  perception of naturalistic stimuli}. In \bibinfo{booktitle}{\emph{Proceedings
  of the 6th Symposium on Applied Perception in Graphics and Visualization}}.
  \bibinfo{pages}{15--18}.
\newblock


\bibitem[\protect\citeauthoryear{Spittle, Kremer, and Hamilton}{Spittle
  et~al\mbox{.}}{2010}]%
        {spittle2010effect}
\bibfield{author}{\bibinfo{person}{Michael Spittle}, \bibinfo{person}{Peter
  Kremer}, {and} \bibinfo{person}{Justin Hamilton}.}
  \bibinfo{year}{2010}\natexlab{}.
\newblock \showarticletitle{The effect of screen size on video-based perceptual
  decision making tasks in sport}.
\newblock \bibinfo{journal}{\emph{International Journal of Sport and Exercise
  Psychology}} \bibinfo{volume}{8}, \bibinfo{number}{4} (\bibinfo{year}{2010}),
  \bibinfo{pages}{360--372}.
\newblock


\bibitem[\protect\citeauthoryear{Spjut}{Spjut}{2022}]%
        {spjut22smalltargets}
\bibfield{author}{\bibinfo{person}{Josef Spjut}.}
  \bibinfo{year}{2022}\natexlab{}.
\newblock \bibinfo{title}{{I}mproving {A}iming {T}ime on {S}mall {F}{P}{S}
  {T}argets with {H}igher {R}esolutions and {L}arger {S}creen {S}izes |
  {N}{V}{I}{D}{I}{A} {T}echnical {B}log --- developer.nvidia.com}.
\newblock
  \bibinfo{howpublished}{\url{https://developer.nvidia.com/blog/improving-aiming-time-on-small-fps-targets-with-higher-resolutions-and-larger-screen-sizes/}}.
\newblock
\newblock
\shownote{[Accessed 09-May-2023].}


\bibitem[\protect\citeauthoryear{Spjut, Boudaoud, Binaee, Kim, Majercik,
  McGuire, Luebke, and Kim}{Spjut et~al\mbox{.}}{2019a}]%
        {spjut2019latency}
\bibfield{author}{\bibinfo{person}{Josef Spjut}, \bibinfo{person}{Ben
  Boudaoud}, \bibinfo{person}{Kamran Binaee}, \bibinfo{person}{Jonghyun Kim},
  \bibinfo{person}{Alexander Majercik}, \bibinfo{person}{Morgan McGuire},
  \bibinfo{person}{David Luebke}, {and} \bibinfo{person}{Joohwan Kim}.}
  \bibinfo{year}{2019}\natexlab{a}.
\newblock \showarticletitle{Latency of 30 Ms Benefits First Person Targeting
  Tasks More Than Refresh Rate Above 60 Hz}. In
  \bibinfo{booktitle}{\emph{SIGGRAPH Asia 2019 Technical Briefs}} (Brisbane,
  QLD, Australia) \emph{(\bibinfo{series}{SA '19})}.
  \bibinfo{publisher}{Association for Computing Machinery},
  \bibinfo{address}{New York, NY, USA}, \bibinfo{pages}{110–113}.
\newblock
\showISBNx{9781450369459}
\urldef\tempurl%
\url{https://doi.org/10.1145/3355088.3365170}
\showDOI{\tempurl}


\bibitem[\protect\citeauthoryear{Spjut, Boudaoud, Binaee, Majercik, McGuire,
  and Kim}{Spjut et~al\mbox{.}}{2019b}]%
        {Spjut19FPSci}
\bibfield{author}{\bibinfo{person}{Josef Spjut}, \bibinfo{person}{Ben
  Boudaoud}, \bibinfo{person}{Kamran Binaee}, \bibinfo{person}{Alexander
  Majercik}, \bibinfo{person}{Morgan McGuire}, {and} \bibinfo{person}{Joohwan
  Kim}.} \bibinfo{year}{2019}\natexlab{b}.
\newblock \showarticletitle{FirstPersonScience: Quantifying Psychophysics for
  First Person Shooter Tasks}. In \bibinfo{booktitle}{\emph{UCI Esports
  Conference}}. \bibinfo{publisher}{UCI}, \bibinfo{address}{Irvine, CA},
  \bibinfo{numpages}{7}~pages.
\newblock


\bibitem[\protect\citeauthoryear{Tan, Gergle, Scupelli, and Pausch}{Tan
  et~al\mbox{.}}{2006}]%
        {tan2006physically}
\bibfield{author}{\bibinfo{person}{Desney~S. Tan}, \bibinfo{person}{Darren
  Gergle}, \bibinfo{person}{Peter Scupelli}, {and} \bibinfo{person}{Randy
  Pausch}.} \bibinfo{year}{2006}\natexlab{}.
\newblock \showarticletitle{Physically large displays improve performance on
  spatial tasks}.
\newblock \bibinfo{journal}{\emph{ACM Transactions on Computer-Human
  Interaction (TOCHI)}} \bibinfo{volume}{13}, \bibinfo{number}{1}
  (\bibinfo{year}{2006}), \bibinfo{pages}{71--99}.
\newblock


\bibitem[\protect\citeauthoryear{Wang, Yu, Qin, Li, and Shi}{Wang
  et~al\mbox{.}}{2013}]%
        {wang2013exploring}
\bibfield{author}{\bibinfo{person}{Yuntao Wang}, \bibinfo{person}{Chun Yu},
  \bibinfo{person}{Yongqiang Qin}, \bibinfo{person}{Dan Li}, {and}
  \bibinfo{person}{Yuanchun Shi}.} \bibinfo{year}{2013}\natexlab{}.
\newblock \showarticletitle{Exploring the effect of display size on pointing
  performance}. In \bibinfo{booktitle}{\emph{Proceedings of the 2013 ACM
  international conference on Interactive tabletops and surfaces}}.
  \bibinfo{pages}{389--392}.
\newblock


\bibitem[\protect\citeauthoryear{Watson, Shrivastava, and Gavane}{Watson
  et~al\mbox{.}}{2019}]%
        {watson2019effects}
\bibfield{author}{\bibinfo{person}{Benjamin Watson}, \bibinfo{person}{Rachit
  Shrivastava}, {and} \bibinfo{person}{Ajinkya Gavane}.}
  \bibinfo{year}{2019}\natexlab{}.
\newblock \showarticletitle{The Effects of Adaptive Synchronization on
  Performance and Experience in Gameplay}.
\newblock \bibinfo{journal}{\emph{Proceedings of the ACM on Computer Graphics
  and Interactive Techniques}} \bibinfo{volume}{2}, \bibinfo{number}{1}
  (\bibinfo{year}{2019}), \bibinfo{pages}{1--13}.
\newblock


\end{thebibliography}

\end{document}